\documentclass{eas}
\usepackage{graphicx}
\begin{document}

\title{Molecular line probes of activity in galaxies} 
\author{S.~Garc\'{\i}a-Burillo}\address{Observatorio de Madrid, OAN, Alfonso XII, 3, 28014-Madrid, SPAIN\email{s.gburillo@oan.es}}
\author{J.~Graci\'a-Carpio}\sameaddress{1}
\author{A.~Usero}\sameaddress{1}
\author{P.~Planesas}\sameaddress{1}
\author{A.~Fuente}\sameaddress{1}
\author{M.~Krips}\address{CfA, SMA project, 60 Garden Street, MS 78 Cambridge, MA 02138, USA}
\begin{abstract}
The use of specific tracers of the dense molecular gas phase can help to explore the feedback of activity on the interstellar medium (ISM) in galaxies. This information is a key to any quantitative assessment of the efficiency of the star formation process in galaxies. We present the results of a survey devoted to probe the feedback of activity through the study of the excitation and chemistry of the dense molecular gas in a sample of local universe starbursts and active galactic nuclei (AGNs). Our sample includes also 17 luminous and ultraluminous infrared galaxies (LIRGs and ULIRGs). From the analysis of the LIRGs/ULIRGs subsample, published in Graci\'a-Carpio \etal\ (\cite{gr07}), we find the first clear observational evidence that the star formation efficiency of the dense gas, measured by the L$_{FIR}$/L$_{HCN}$  ratio, is significantly higher in LIRGs and ULIRGs than in normal galaxies. Mounting evidence of overabundant HCN in active environments would even reinforce the reported trend, pointing to a significant turn upward in the Kennicutt-Schmidt law around L$_{FIR}$=10$^{11}$L$_{\odot}$. This result has major implications for the use of HCN as a tracer of the dense gas in local and high-redshift luminous infrared galaxies.
\end{abstract}
\maketitle
\section{Tracking down galaxy evolution through chemistry}
Millimeter telescopes can provide a detailed picture of the distribution and kinematics of molecular gas in galaxy disks through extensive CO line mapping. However, the use of tracers more specific to the dense molecular gas phase, the one directly involved in the fueling of star formation and AGN episodes, is required to probe the feedback of activity on the ISM. Provided that high spatial resolution and high sensitivity requirements are met, the observation of complex molecular species can help to track down galaxy evolution. Extragalactic chemistry can be used to challenge current chemical models of molecular gas, as active galaxies can drive chemical complexity on much larger scales compared to our Galaxy. Furthermore, this type of studies are a key to constrain conversion factors, which are required to derive gas masses from line luminosities. This is mostly relevant, as the excitation and chemistry of some of the more routinely used dense gas tracers (e.g., HCN lines) can be heavily affected in active environments.   
 
 Different processes can shape the evolution of molecular gas along the typical evolutionary track of a starburst: large-scale shocks, strong UV-fields, cosmic-rays, and eventually X-rays, in the presence of an AGN. These actors are expected to be at play at different stages but also at different locations in the disk and in the disk-halo interface of galaxies showing activity. The use of interferometers makes possible to unveil a strong chemical differentiation in the molecular gas disks of active galaxies. The example of M82, a prototypical starburst, is paradigmatic in this respect. Virtually all of the large-scale SiO emission detected in the PdBI map of M~82, published by Garc\'{\i}a-Burillo \etal\ (\cite{gb01}), traces the disk-halo interface of the galaxy where episodes of mass injection are building up the gaseous halo. In contrast, widespread emission of the formyl radical, HCO, mapped in M~82 with the PdBI, reveals the propagation of photo-dissociation region (PDR) chemistry inside the disk of this starburst (Garc\'{\i}a-Burillo \etal\ \cite{gb02}). The scenario of a giant PDR in the disk of M~82 has received further support from the high abundances measured for molecular tracers specific to PDR such as CN, HOC$^+$, CO$^+$ as well as a set of small hydrocarbon chains (Fuente \etal\ \cite{fu05,fu06}).

\section{Molecular gas chemistry in AGNs}

\begin{figure}
\begin{center}
\includegraphics[width=11.5cm]{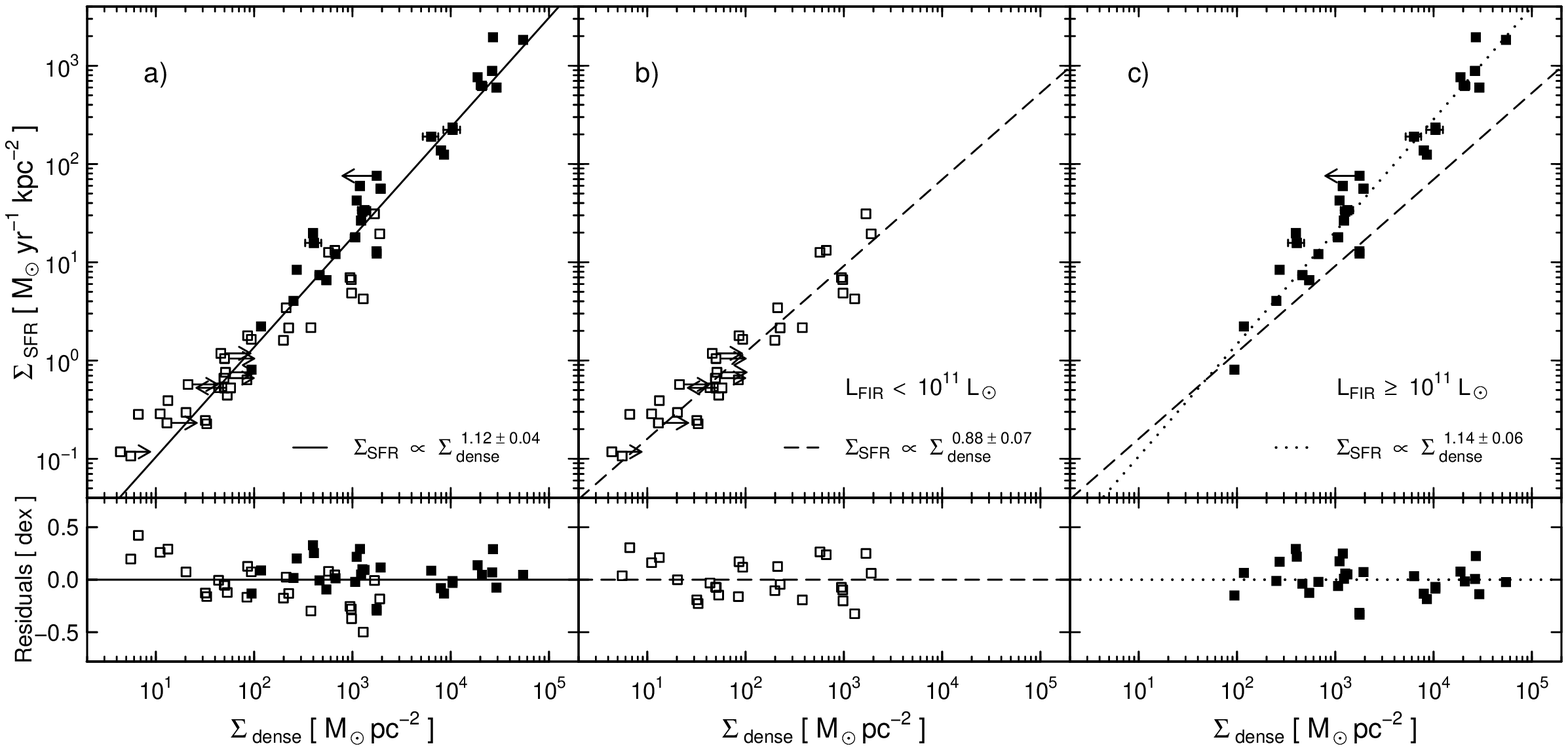}
\caption{{\bf (a)} Surface density of star formation rate, $\Sigma_{\rm SFR}$, against surface density of the dense molecular gas mass derived from HCN, $\Sigma_{\rm dense}$, for normal galaxies (open squares) as well as LIRGs, ULIRGs and high-z objects (filled squares). The solid line is the regression fit to the full sample of galaxies. In {\bf (b)} and {\bf (c)} we separate between normal and LIRGs/ULIRGs/high-z, respectively. The dashed (dotted) line is the fit to galaxies with $L_{\rm FIR} < 10^{11} L_{\odot}$ ($L_{\rm FIR} > 10^{11} L_{\odot}$). Lower panels show the residuals of the fits. (Adapted from Graci\'a-Carpio \etal\ \cite{gr07}).}
\end{center}
\end{figure}
There is increasing evidence that the excitation and the chemistry of molecular gas show significant differences between  starbursts and AGNs (e.g., Kohno \etal\ \cite{ko01}; Krips \etal\ \cite{kr07}). In particular, Krips \etal\ (\cite{kr07}) have recently shown that starbursts and AGNs populate different regions in a set of diagnostic diagrams 
that make use of HCN and HCO$^+$ line ratios. LVG fits to the observed ratios point to lower H$_2$ densities and larger HCN abundances in AGNs compared to starburst galaxies. Overluminous HCN lines (possibly leading to overabundant HCN) seem to be the rule in the circumnuclear disks (CND) of many Seyferts. The 200~pc-radius CND of NGC~1068 is viewed as the first clear example of a giant X-ray dominated region (XDR) leading to enhanced abundances of some molecular species like HCN (Usero \etal\ \cite{us04}). Recently a high-resolution PdBI map has shown that strong SiO emission comes from the CND of NGC~1068 (Garc\'{\i}a-Burillo \etal\ 2008, in prep.). The enhancement of SiO in this case cannot be attributed to the action of ongoing star formation as there is negative evidence of a recent starburst in the CND of NGC~1068 (Davies \etal\ \cite{da07}). Alternatively, the processing of 10~\AA~dust grains by X-rays, as a mechanism to enhance silicon chemistry in gas phase, would explain the large SiO abundances of the CND.
In particular, the inclusion of dust grain chemistry could solve the controversy between different gas-phase XDR schemes as concerns the abundance of some molecular tracers like HCN in X-ray irradiated environments. 
\begin{figure}
\begin{center}
\includegraphics[width=10cm]{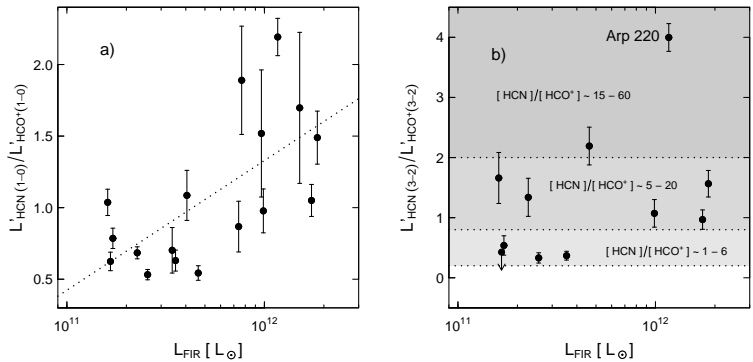}
\caption{{\bf (a)} HCN(1--0)$/$HCO$^{+}$(1--0) luminosity ratio as a function of $L_{\rm FIR}$ in the sample of LIRGs and ULIRGs of Graci\'a-Carpio \etal\ (\cite{gr07}). The linear fit to the data (dashed line) shows the trend. {\bf (b)} Same as {\bf (a)} but for the J=3--2 line.
We indicate the [HCN]/[HCO$^+$] abundance ratios derived from multiline LVG analysis.}
\end{center}
\end{figure}
\section{Molecular gas chemistry in IR luminous galaxies}

The caveats on the use of HCN as a tracer of dense gas in active galaxies call for the use of alternative tracers of dense gas in LIRGs and ULIRGs, galaxies where star formation and AGN activity are expected to be highly embedded. This question is paramount to disentangle the power sources of the infrared luminosities of these galaxies. In a recent paper, Graci\'a-Carpio \etal\ (\cite{gr07}) have presented evidence that the $L_{\rm FIR}/L_{\rm HCN(1-0)}$ ratio, taken as a fair proxy for the star formation efficiency of the dense gas (SFE$_{\rm dense}$), is a factor $\sim$2--3 higher in IR luminous galaxies ($L_{\rm FIR} > 10^{11}\,L_{\odot}$) compared to normal galaxies. Local universe LIRGs and ULIRGs populate a region in the SFE$_{\rm dense}$ diagram that lies between those occupied by normal and high-$z$ IR luminous galaxies. The reported trend in the SFE$_{\rm dense}$ derived from HCN data implies that there is a statistically significant turn upward in the Kennicutt-Schmidt law, $\Sigma_{\rm SFR} \propto \Sigma_{\rm dense}^{N}$, at high $L_{\rm FIR}$: $N$ changes from $\sim$0.80--0.95 (for $L_{\rm FIR} < 10^{11}\,L_{\odot}$) to $\sim$1.1--1.2 (for $L_{\rm FIR} > 10^{11}\,L_{\odot}$) (See Fig.~1).
 
Based on a multiline LVG analysis of HCN and HCO$^{+}$ data, that follows the survey published by Graci\'a-Carpio \etal\ (\cite{gr06}), Graci\'a-Carpio \etal\ (\cite{gr07}) find that the the conversion factor between $L_{\rm HCN(1-0)}$ and M$_{dense}$, $X_{\rm HCN}$, is $\sim$3 times lower at high $L_{\rm FIR}$ (See Fig.~2). Of particular note, a significant overabundance of HCN has also been reported in the IR luminous z$\sim$4 quasar APM~08279 (Wagg \etal\ \cite{wa05}; Garc{\'{\i}}a-Burillo \etal\ \cite{gb06}). This reinforces the scenario where the SFE$_{\rm dense}$ could well be an order of magnitude higher in extreme LIRGs/ULIRGs compared to normal galaxies. 

%
%

%

%

\end{document}